\documentclass[%
 reprint,
nofootinbib,
nobibnotes,
 amsmath,amssymb,
 aps, prd,
 longbibliography,
floatfix,
]{revtex4-2}

\usepackage{graphicx}
\usepackage{dcolumn}
\usepackage{bm}
\usepackage[normalem]{ulem}

\usepackage[english]{babel}
\usepackage{fancyhdr}
\usepackage{amsmath}
\usepackage{amssymb}
\usepackage{amsfonts}
\usepackage{psfrag}
\usepackage[utf8]{inputenc}
\usepackage[bf,footnotesize,justification=Justified,format=plain]{caption}
\usepackage[dvipsnames]{xcolor}
\usepackage[colorlinks=True, citecolor=blue, linkcolor=blue, urlcolor=blue,linktocpage]{hyperref}
\usepackage{amsthm}
\usepackage{gensymb}
\usepackage{subfig}
\usepackage{physics}
\usepackage{array}
\usepackage{tcolorbox, mathtools}
\usepackage{soul}
\tcbuselibrary{skins}
\newtcolorbox{mybox}[1][]{before=\centering, drop fuzzy shadow, enhanced, colframe=blue, fonttitle=\bfseries, title=#1, center title}
\usepackage{aas_macros}
\usepackage{cleveref}

\newcommand{\sch}{Schwarzschild }

\newcommand\underrel[2]{\mathrel{\mathop{#2}\limits_{#1}}}

\newcommand{\centra}{CENTRA, Departamento de Física, Instituto Superior Técnico – IST,
Universidade de Lisboa – UL, Avenida Rovisco Pais 1, 1049-001 Lisboa, Portugal}

\begin{document}

\title{Green function of the P\"{o}schl-Teller potential}

\author{Adrien Kuntz}
 \email{adrien.kuntz@tecnico.ulisboa.pt}
\affiliation{\centra}


\begin{abstract}
We use an approximation of the Regge-Wheeler-Zerilli potential, known as P\"{o}schl-Teller, to exactly compute the time-domain Green function of black hole perturbations in this simplified model, taking into account all causality conditions. We find the existence of an additional \textit{early times} piece in the Green function, contributing to new exponentially growing modes just before the signal interacts with the maximum of the potential. The waveform itself is decomposed as an instantaneous piece traveling exactly on the light-cones of the Green function and a historical piece depending on the past trajectory of the system inside the light-cone. We also study redshift modes and show that the Regge-Wheeler-Zerilli Green function is regular at their frequency, with no zero nor pole.
\end{abstract}

\date{\today}

\maketitle

\section{Introduction}

Solving Einstein's equations as accurately as possible for a merger of two black holes (BH) is a crucial step in order to understand the nature of these fascinating astronomical objects. Right after the merger, our current understanding of waveforms points towards the existence of Quasi-Normal Modes (QNM) oscillations~\cite{Berti:2025hly}. In this picture, each harmonic of the waveform can be decomposed in an infinite sum of damped sinusoids, whose frequencies and damping times only depend on the final mass and spin of the BH.

However, the full content of the waveform is usually richer than the simple linear QNM picture. At late times, power-law tails dominate the signal~\cite{Price:1972pw,PhysRevD.34.384,PhysRevD.55.468,DeAmicis:2024not, DeAmicis:2024eoy}. Nonlinearities can manifest themselves though the emergence of additional quadratic QNM frequencies~\cite{Mitman:2022qdl, Cheung:2022rbm, Bucciotti:2023ets, Bucciotti:2024zyp, Bucciotti:2024jrv, Khera:2024bjs, Yi:2024elj, Lagos:2024ekd}. Around the merger time, it is still unclear what is the contribution of the QNMs to the total waveform, since there can be other transient signals which are important as well~\cite{Leaver:1986gd,Lagos_2023, Chavda:2024awq,PhysRevD.55.468}.

This last issue will be the central theme of the article. We will focus on the plunge of a small BH, represented as a point-particle, into a larger one. In this case, one can use the tools from BH perturbation theory to describe the signal emitted by the smaller BH as it moves along its plunge trajectory. While it is always possible to solve these equations numerically, physical insight comes in when introducing the Green function of the BH perturbation equations. The seminal article from Leaver~\cite{Leaver:1986gd} identified the QNM contribution to the waveform to a series of poles of the Green function in the complex plane. The full Green function itself was decomposed by Leaver into three pieces: QNM, tail and prompt response. However, the impact of the tail and prompt response on the waveform at early times, i.e. close to the merger, was still left as an open issue in this article.

This problem is still open today, mainly due to the fact that there is no known closed-form expression for the time-domain Green function which could allow to neatly separate the signal into different pieces and evaluate each one of them separately. In this article, we propose to bypass this difficulty by computing exactly the Green function for a different kind of potential in the Regge-Wheeler-Zerilli (RWZ) equations obeyed by perturbations of a nonspinning BH. This P\"{o}schl-Teller potential~\cite{Poschl:1933zz,Ferrari:1984zz,Beyer:1998nu} is tuned to ``look like'' the RWZ potential close to its maximum, where we expect that the dominant part of the signal is produced close to merger. The P\"{o}schl-Teller potential has been used in the past to give insight on the computation of QNM~\cite{Ferrari:1984zz,Beyer:1998nu}, since the wave equation in this potential can be solved exactly in terms of hypergeometric functions. It has also been used to understand the structure of the QNM part of the Green function~\cite{Casals:2009zh, Casals:2013mpa}. We thus expect that it should also give physical intuition about the decomposition of the waveform close to merger.

As it should be clear, the aim of this article is to concentrate on physical intuition, not on precision. Indeed, the results that we will obtain in Section~\ref{sec:response} are a poor fit to the true waveforms of plunging BHs. However, we will find several surprising new results showing that the Leaver decomposition into QNM, tail and prompt response is probably not a true representation of the signal at all times. Indeed, even in this simplified model, we find that the prompt response (integral of the Green function along high-frequency arcs) is zero, while there is another \textit{early time} contribution to the Green function at times before the onset of QNM ringing. This contribution comes from a decomposition of the Green function into ingoing and outgoing components before QNM ringing. 
In some sense, our results can be considered as an improvement of the toy-models considered in~\cite{Lagos_2023,Chavda:2024awq,Szpak:2004sf}, where the potential was a simpler delta-function or a decaying exponential. They can also be regarded as a quantification of how much the branch cut integral contributes to the Green function of the RWZ potential at all times, since the P\"{o}schl-Teller potential can be regarded as a RWZ potential to which one has removed the power-law tail, and the branch cut is known to originate from the long-range properties of the potential~\cite{Leaver:1986gd, PhysRevD.55.468}.

Going further and using our analytic expression for the Green function to integrate the equations for a plunging point-particle, we will find that the resulting waveform can be decomposed in several pieces. The first one is a \textit{direct propagation} contribution, travelling exactly on the light-cones of the Green function~\footnote{Notice that, in this article, we will loosely denote as ``light-cones of the Green function'' the characteristic lines of the Green function for the $\ell=2$ gravitational radiation. They are different from the light-cones of the full \sch Green function, which are null geodesics of the \sch spacetime. The latter are recovered by summing the multipole-expanded Green function over all the different $\ell$ modes, see e.g.~\cite{Casals:2019heg}. }. This piece just tracks the source term of the RWZ equations at retarded times. The second piece is \textit{historical}, i.e. it depends on the past history of the system inside the light cone. QNM oscillations are part of this historical wavefunction, but there is also an early time historical piece. 

We note that a recent article~\cite{DeAmicis:2025xuh} has computed the historical QNM piece of the full RWZ equations, properly taking into account causality conditions. When possible, we will qualitatively compare our results to theirs, showing that we also recover the so-called \textit{redshift} (or \textit{horizon}) modes in the waveform in the regime where it is usually thought that only QNM oscillations subsist. The existence of these new modes could sensibly modify our picture of the waveform in the QNM regime, as the dominant redshift mode is less damped than the first overtone for $\ell=2$ perturbations. On the other hand, we will explain why we disagree with a recent proposal indicating a screening of the redshift modes~\cite{Oshita:2025qmn} in Section~\ref{sec:remarks}.

This paper is organized as follows. In Section~\ref{sec:defGreen}, we will recap how one can solve BH perturbation equations using the Green function technique; this part can be skipped by the readers already familiar with the topic. In Section~\ref{sec:PTGreen} we introduce the P\"{o}schl-Teller potential and exactly solve its Green function at all times using an integration contour in the complex plane. We also compare our analytical results to a numerical solution, finding an excellent agreement. Next, in Section~\ref{sec:response} we compute the waveform generated by a point-particle plunging into the BH along two different trajectories. Finally, in Section~\ref{sec:remarks} we make some remarks concerning the analytic structure of the full RWZ equations Green function. The aim of this last section is to highlight a discrepancy with Ref~\cite{Oshita:2025qmn} as well as to provide a qualitative discussion on how our results generalize to \sch. 
We use units in which $G=1$. Furthermore we set the BH mass to be $2M=1$, so that the horizon is located at $r=1$ in \sch coordinates.

\section{Defining the Green function} \label{sec:defGreen}

Linear perturbations $\psi(x,t)$ of the \sch geometry obey the Regge-Wheeler-Zerilli equations~\cite{Regge:1957td, Zerilli:1970aa},
\begin{equation}
   - \frac{\partial^2 \psi}{\partial t^2} + \frac{\partial^2 \psi}{\partial x^2} - V \psi = S \label{eq:RWZeq} \; ,
\end{equation}
where $x$ is the tortoise coordinate related to \sch radial distance $r$ by $x = r + \log(r-1)$, $V(x)$ is the Regge-Wheeler (odd) or Zerilli (even) potential, and $S$ is an arbitrary source term (we will consider $S$ to be the source term corresponding to a point-particle orbiting the BH in Section~\ref{sec:response}). All our calculations will be carried out with the Zerilli potential for $\ell=2$. Notice that we do not aim to discuss here the full 4-dimensional structure of the Green function but we restrict to $(t,r)$ space: the caustic structure, light crossing time and geodesics are known to be different in the two cases~\cite{Casals:2009zh}.
In this section, we will briefly recap how one can solve Eq.~\eqref{eq:RWZeq} subject to the free radiation boundary conditions (i.e. $\psi$ is an ingoing wave at the horizon and an outgoing wave at infinity) using the Green function technique.

In order to decompose Eq.~\eqref{eq:RWZeq} in the frequency domain, we introduce the Laplace transform of $\psi$,
\begin{equation}
    \tilde \psi(x, \omega) = \int_{t_0}^\infty \psi(x,t) e^{i \omega t} \dd t \; .
\end{equation}
Here, $t_0$ is the initial time at which we impose initial conditions $\psi_0 = \psi(x,t_0)$ and $\dot \psi_0 = \partial \psi(x,t)/\partial t |_{t_0}$. Mathematically, the initial time is required in the integral since it could otherwise diverge for complex values of $\omega$. The Laplace transform is usually an analytic function of $\omega$ for sufficiently large values of $\mathrm{Im} \, \omega$, while it can have poles and branch cuts in the rest of the complex plane. It can be inverted with the formula
\begin{equation}
    \psi(x,t) = \int_{i \gamma - \infty}^{i \gamma+\infty} \tilde \psi(x,\omega) e^{-i \omega t} \frac{\dd \omega}{2 \pi} \label{eq:inverseLaplaceTransform} \; .
\end{equation}
Here, $\gamma$ is a real number chosen so that the line integral in~\eqref{eq:inverseLaplaceTransform} lies within the portion of complex plane where $\tilde \psi$ is analytic. For our applications, we will choose $\gamma = 0^+$ to be a small positive number.

Applying the Laplace transform to Eq.~\eqref{eq:RWZeq}, we find the equation in the frequency domain,
\begin{equation}
    \frac{\partial^2 \tilde \psi}{\partial x^2} + (\omega^2-V) \tilde \psi = \tilde S - e^{i \omega t_0} \dot \psi_0 + i \omega e^{i \omega t_0} \psi_0 \label{eq:RWZeqFreqDomain} \equiv J(x, \omega) \; ,
\end{equation}
where $\tilde S$ is the Laplace transform of $S$. Here, the additional terms on the right-hand side serve as a source representing initial conditions. 
We can solve this equation with the variation of constants (or Green function) method. Let us introduce two solutions $\psi_H^-(x, \omega)$ and $\psi_\infty^+(x, \omega)$ of the homogeneous equation~\eqref{eq:RWZeqFreqDomain} normalized such that $\psi_H^-$ represents an ingoing wave at the horizon and  $\psi_\infty^+$ represents an outgoing wave at infinity,
\begin{equation} \label{eq:largeXAsymptotics}
    \psi_H^- (x, \omega)\underrel{x \rightarrow - \infty}{\simeq} e^{-i \omega x} \; , \quad  \psi_\infty^+ (x, \omega)\underrel{x \rightarrow  \infty}{\simeq} e^{i \omega x} \; ,
\end{equation}
and we define their constant Wronskian by $W(\omega) = \psi_H^- (\psi_\infty^+)' - (\psi_H^-)' \psi_\infty^+$, where $'=\dd/\dd x$. In the next sections of the paper, we will also need a third homogeneous solution which is purely \textit{ingoing} at infinity $\psi_\infty^- \underrel{x \rightarrow  \infty}{\simeq} e^{-i \omega x}$. 
Then, we define the Green function by
\begin{align}
    \tilde G(x, \bar x, \omega) &= \frac{\Theta(\bar x - x)}{W} \psi_\infty^+(\bar x, \omega) \psi_H^-(x, \omega) \nonumber \\ 
    &+ \frac{\Theta( x - \bar x)}{W} \psi_H^-(\bar x, \omega) \psi_\infty^+( x, \omega) \; , \label{eq:defGreen}
\end{align}
where $\Theta$ is the Heaviside function. The Green function solves the differential equation
\begin{equation}
    \frac{\partial^2 \tilde G}{\partial x^2} + (\omega^2-V) \tilde G = \delta(x-\bar x) \; ,
\end{equation}
and hence the solution $\tilde \psi$ to Eq.~\eqref{eq:RWZeqFreqDomain} subject to the free radiation boundary condition is given by
\begin{equation}
    \tilde \psi(x,\omega) = \int_{-\infty}^\infty \dd \bar x \tilde G(x, \bar x, \omega) J(\bar x, \omega) \label{eq:solTildePsi} \; .
\end{equation}
We now define the time-domain Green function $G(x, \bar x, t)$ just by inverse Laplace transform,
\begin{equation}
    G(x,\bar x,t) =  \int_{i \gamma - \infty}^{i \gamma+\infty} \tilde G(x, \bar x,\omega) e^{-i \omega t} \frac{\dd \omega}{2 \pi} \label{eq:InverseLaplaceGree} \; .
\end{equation}
Physically, it can be identified with the signal that reaches an observer at time $t$ at position $x$ for a small localized perturbation emitted at $t=0$ at position $\bar x$. 
Using the solution for $\tilde \psi$ in Eq.~\eqref{eq:solTildePsi} with the definition of the source in Eq.~\eqref{eq:RWZeqFreqDomain}, it is easy to show that the time-domain field $\psi$ is given by
\begin{align} \label{eq:LeaverStrangeFormula}
    &\psi(x, t) = \int_{- \infty}^\infty \mathrm{d}\bar x \int_{t_0}^\infty \dd \bar t \; G(x, \bar x, t-\bar t) S(\bar x, \bar t) \nonumber \\
    &- \int_{- \infty}^\infty \mathrm{d}\bar x \big[ G(x, \bar x, t-t_0) \dot \psi_0(\bar x) + \partial_t G(x, \bar x, t-t_0) \psi_0(\bar x) \big] \; .
\end{align}
We have thus reduced the problem to finding an inverse Laplace transform for the Green function in Eq.~\eqref{eq:InverseLaplaceGree}. This is the topic of the next Section.

\section{The P\"{o}schl-Teller Green function} \label{sec:PTGreen}

\begin{figure}
    \centering
    \includegraphics[width=\linewidth]{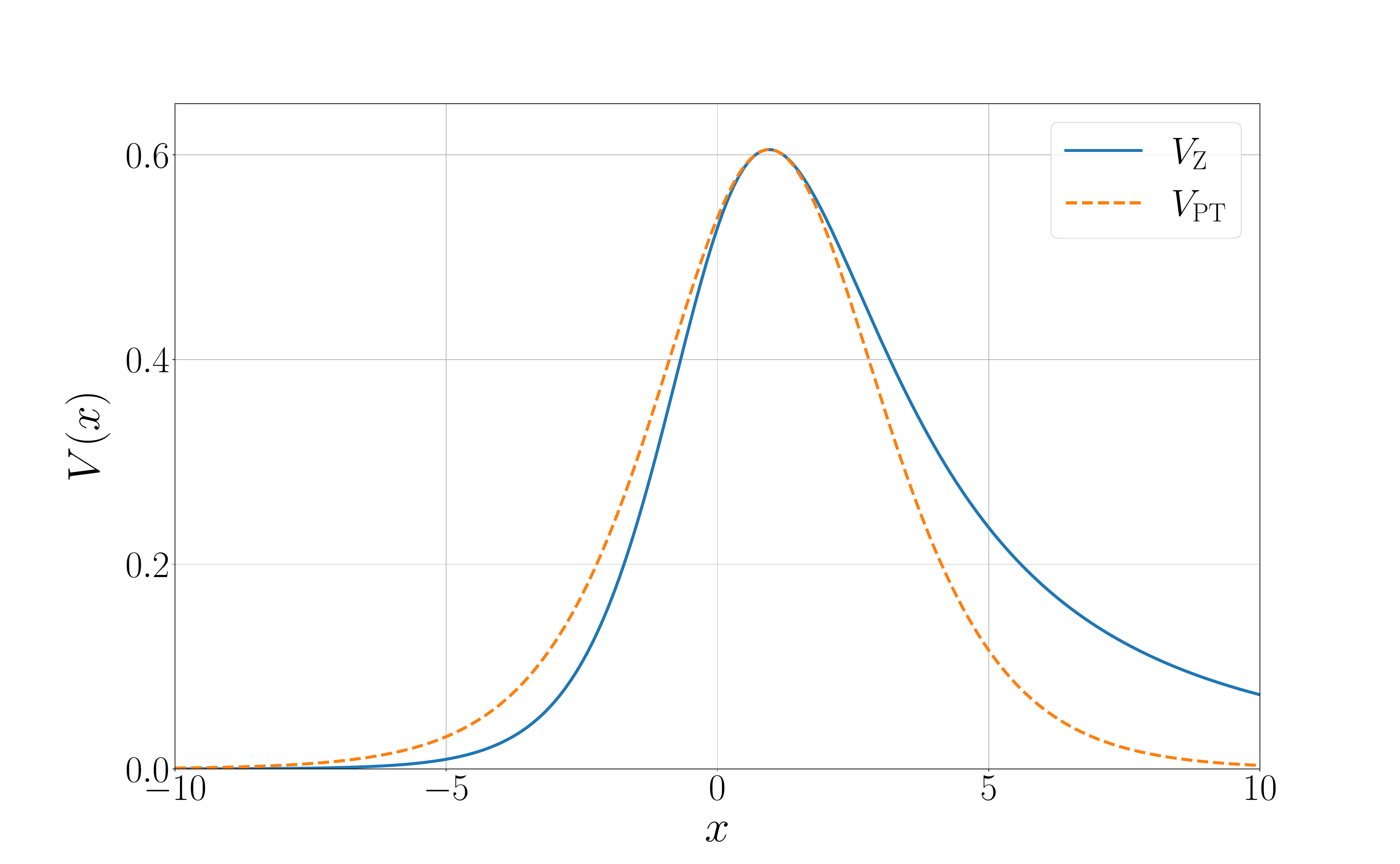}
    \caption{Comparison of the Zerilli potential $V_\mathrm{Z}$ for $\ell=2$ and the P\"{o}schl-Teller potential $V_\mathrm{PT}$ as a function of the tortoise coordinate $x$. }
    \label{fig:potential_comparison}
\end{figure}

Inverting the Laplace transform in Eq.~\eqref{eq:InverseLaplaceGree} is notoriously difficult due to the complicated analytical properties of the solutions of the RWZ equations, known as Heun functions~\cite{Fiziev:2005ki}. The main idea of this article is to solve the Green function for another potential which ``looks like" the RWZ potential close to its maximum, and for which the Green function can be obtained exactly in terms of hypergeometric functions. This is the P\"{o}schl-Teller potential~\cite{Poschl:1933zz, Ferrari:1984zz, Beyer:1998nu, Berti:2009kk}, given by
\begin{equation}
    V_{PT}(x) = \frac{V_0}{\mathrm{cosh}^2 \alpha(x - x_m)} \; .
\end{equation}
Here, $x_m$ is the location of the maximum, $V_0 = V(x_m)$ represents the height of the potential at its maximum and $\alpha$ is related to the second derivative of $V$ at its maximum, $\alpha^2 = - \frac{\mathrm{d}^2 V}{\mathrm{d}x^2} \big|_{x_m} /(2V_0)$. We plot in Figure~\ref{fig:potential_comparison} the P\"{o}schl-Teller potential on top of the Zerilli potential for $\ell=2$, with $x_m$, $V_0$ and $\alpha$ tuned to fit the maximum of both potentials using their definition. Specifically, $x_m = 0.95$, $V_0 = 0.605$ and $\alpha = 0.362$. 

Our hope is that the Green function of the  P\"{o}schl-Teller potential will be ``close" enough to the one of the RWZ potential, at least in the vicinity of the maximum. 
It is well-known that great care should be exercised when comparing the properties of solutions to wave equations with similar potentials. Indeed, in the frequency domain, the spectral properties of two closeby potentials can be widely different~\cite{Nollert:1996rf, Cheung:2021bol}. However, it is recognized that the time-domain properties of the wavefunction should be insensitive to small variations in the potential~\cite{Berti:2022xfj}. Thus, we will be mostly interested in the time-domain waveform, where we expect that the  P\"{o}schl-Teller approximation can be good for perturbations located close to the maximum.

With this caveat in mind, we now turn to solving the Green function with the  P\"{o}schl-Teller potential. In order to simplify our problem, in this section we just linearly shift the tortoise coordinate $x \rightarrow x + x_m$ to place the maximum of the potential at $x=0$. 
The solutions to the homogeneous frequency-domain equation~\eqref{eq:RWZeqFreqDomain} are given by~\cite{Berti:2009kk}
\begin{widetext}
\begin{align} 
    \psi_H^-(x, \omega) &= e^{-i \omega x}  F \Big( \lambda, \bar \lambda , 1 - i \omega/\alpha, \big(1+e^{-2 \alpha x} \big)^{-1} \Big) \; , \nonumber \\
    \psi_\infty^-(x, \omega) &= e^{-i \omega x} F \Big(  \lambda, \bar \lambda, 1 + i \omega/\alpha, e^{-2 \alpha x} \big(1+e^{-2 \alpha x} \big)^{-1}  \Big) \; , \nonumber \\
    \psi_\infty^+(x, \omega) &= e^{i \omega x}  F \Big( \lambda , \bar \lambda , 1 - i \omega/\alpha, e^{-2 \alpha x} \big(1+e^{-2 \alpha x} \big)^{-1} \Big) \; , \label{eq:defHomogeneousSolutions}
\end{align}
\end{widetext}
where $F = {}_2F_1$ is the hypergeometric function and the complex parameter $\lambda$ is given by
\begin{equation}
    \lambda = \frac{1}{2} \bigg(1+ i \sqrt{\frac{4 V_0}{\alpha^2}-1} \bigg) \simeq 0.5 + 2.09 i \; .
\end{equation}
Notice that its complex conjugate is given by $\bar \lambda = 1-\lambda$. 
In order to perform the inverse Laplace transform, we have to understand the analytic structure of the Green function defined in Eq.~\eqref{eq:defGreen}. We will be interested in situations where $x>0$ and $x > \bar x$, i.e. for an observer measuring the field far from the source. However, at that stage we do not yet simplify $\psi_\infty^+$ using its asymptotic expansion for large $x$, because asymptotic expansions often lose track of poles and branch cuts in complex functions (we will have more to say on this point later on). Hence, the Green function is given by
\begin{equation} \label{eq:GreenFarObserver}
    \tilde G(x, \bar x, \omega) = \frac{ \psi_H^-(\bar x, \omega) \psi_\infty^+( x, \omega)}{W} \; .
\end{equation}
Of particular importance will be the \textit{connection coefficients} of the ingoing wavefunction $\psi_H^-$:
\begin{equation} \label{eq:defConnectionCoeffs}
    \psi_H^-(x, \omega) = A^-(\omega) \psi_\infty^-(x, \omega) +  A^+(\omega) \psi_\infty^+(x, \omega) \; .
\end{equation}
It is easy to find that the Wronskian is $W = 2 i \omega A^-$. Furthermore, using properties of the hypergeometric function we find
\begin{align} \label{eq:A-}
    A^- &= \frac{\Gamma \big(1-i \omega/\alpha \big) \Gamma \big(-i \omega/\alpha \big)  }{\Gamma \big(\lambda-i \omega/\alpha \big) \Gamma \big(\bar \lambda-i \omega/\alpha \big)} \; , \\
    A^+ &= \frac{2 i \pi}{|\Gamma \big( \lambda \big)|^2 \big( e^{-\pi \omega/\alpha} - e^{\pi \omega/\alpha} \big)} \label{eq:A+} \; .
\end{align}
Hence, the Wronskian $W$ has double poles on the negative imaginary axis $\omega = - i \alpha (k+1)$ with $k \geq 0$ an integer. At the same time, both $ \psi_H^-$ and $ \psi_\infty^+$ have simple poles at these values of $\omega$ (coming from the hypergeometric function), so that the Green function in Eq.~\eqref{eq:GreenFarObserver} is regular at these points. Notice that approximating $\psi_\infty^+ \simeq e^{i \omega x}$ for large $x$ completely loses track of the pole structure of this function, so that we could have incorrectly deduced that $\tilde G$ had zeros on the imaginary axis by doing this approximation. 
On the other hand, the inverse Wronskian $1/W$ has simple poles at the QNM frequencies:
\begin{equation} \label{eq:QNMfreq}
    \omega_n = -i \alpha (n + \lambda) \; , \quad n \in \mathbb{N} \; .
\end{equation}
There is no other pole in the Green function, in particular there is no branch cut contrary to the Green function of the RWZ potential. This can heuristically be understood from the fact that we got rid of the power-law decay of the RWZ potential by replacing it with $V_{PT}$, hence ignoring the tail effect originating from the branch cut. This simplification will allow us to exactly compute the inverse Laplace transform of $\tilde G$.

\begin{figure}
     \centering
     \subfloat[$t < x - \bar x$]{
        \includegraphics[width=0.45\columnwidth]{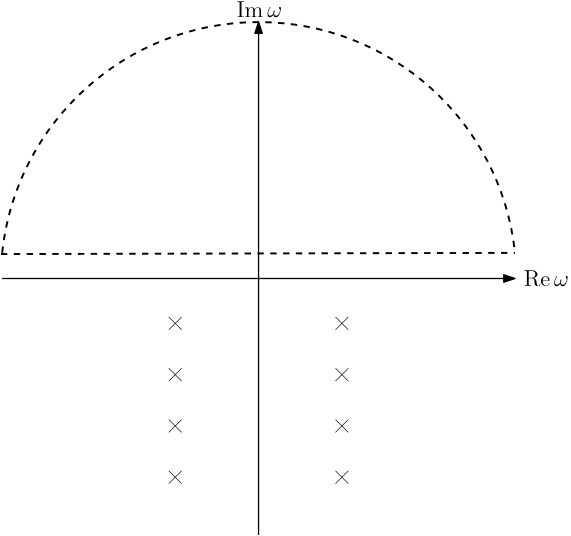}
    }
    \subfloat[$t > x + |\bar x|$]{
        \includegraphics[width=0.45\columnwidth]{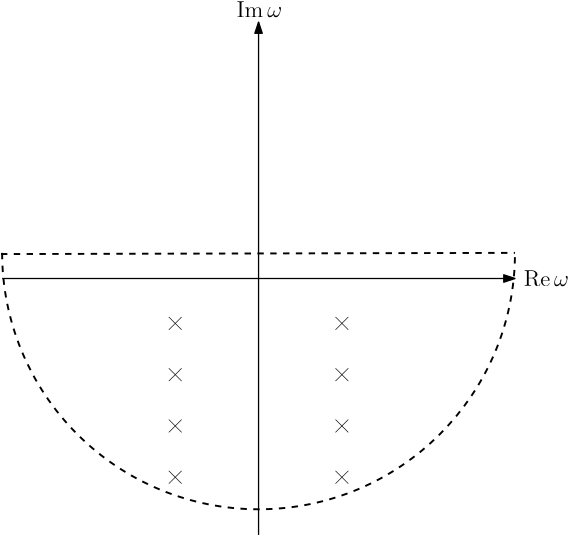}
    } \\
    \subfloat[$x - \bar x < t < x + \bar x$]{
        \includegraphics[width=0.45\columnwidth]{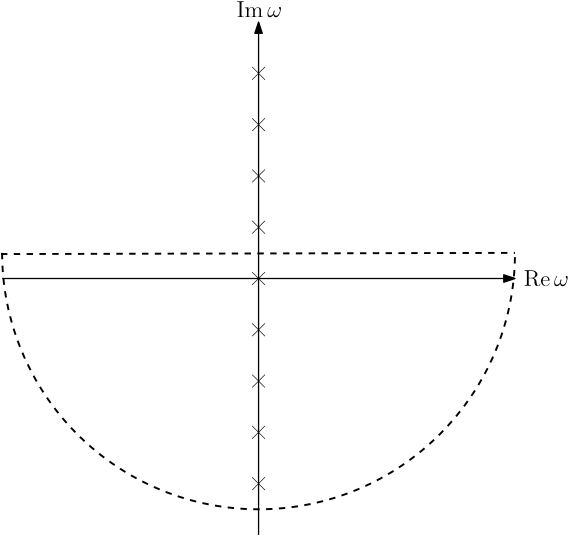}
        \includegraphics[width=0.45\columnwidth]{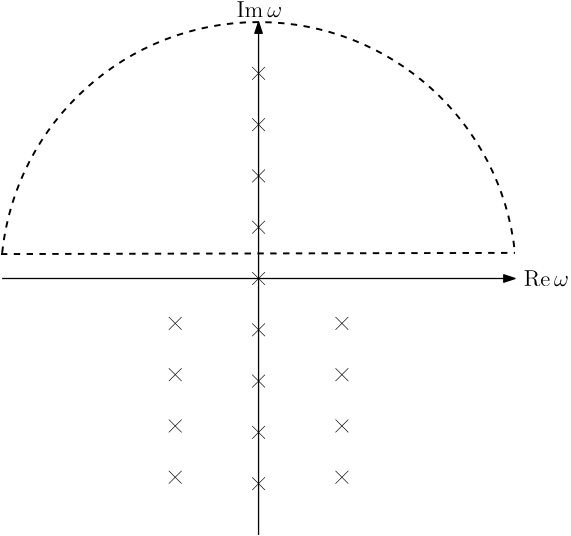}
    }
    \caption{Closing the integration contour according to causality conditions. For $t<x-\bar x$, the contour can be closed in the upper half-plane and the Green function is zero. For $t>x+|\bar x|$ the contour can be closed in the lower half-plane and picks up the residues at QNM frequencies. Finally, for $x-\bar x<t<x+\bar x$ (this only exists provided $\bar x>0$), the Green function is decomposed as a sum of two contours in the upper and lower half-planes picking up residues on the imaginary axis.} \label{fig:contour}
\end{figure}

We use the residue theorem to compute the line integral in~\eqref{eq:InverseLaplaceGree}. We have to decide whether closing the contour on the upper or lower half of the complex plane, making sure that the contribution from the arcs at large $\omega$ is convergent. This depends on the asymptotics of the Green function for large $\omega$. We will split the computation in two different cases, whether $\bar x <0$ or $\bar x >0$, illustrated in Figure~\ref{fig:contour}.

\subsection{First case: $\bar x < 0$} \label{sec:xbar<0}

We use the property that, for large $|\omega|$ and when $z<1/2$, one has $ F \Big( \lambda, \bar \lambda , 1 - i \omega/\alpha,z \Big) \underrel{|\omega| \rightarrow \infty}{\simeq} 1$ as long as $\omega$ does not fall on a pole of the hypergeometric function at $\omega = - i \alpha k$ previously mentioned~\cite{NIST:DLMF}. Thus, for large $|\omega|$
\begin{align}
     \psi_H^-(\bar x, \omega) &\underrel{|\omega| \rightarrow \infty}{\simeq} e^{-i \omega \bar x} \; \text{for} \; \bar x <0 \; , \\
     \psi_\infty^+(x, \omega) &\underrel{|\omega| \rightarrow \infty}{\simeq} e^{i \omega x} \; \text{for} \; x >0  \; .
\end{align}
This is nothing else than the statement that the asymptotics at spatial infinity and at the horizon in Eq.~\eqref{eq:largeXAsymptotics} are in fact valid whatever $\bar x<0$, $x>0$ for large $|\omega|$.

Hence, the Green function behaves as $\tilde G e^{-i \omega t} \sim e^{i \omega (x - \bar x - t)}/W$ along the arcs at large $|\omega|$. Using that $A^- \sim 1$ for large $|\omega|$, we get that the contour can be closed in the upper half plane for $t<x-\bar x$, and in the lower half-plane for $t>x-\bar x$. In both cases the contribution of the arcs are zero by Jordan lemma. 

When closing on the upper half-plane, no pole is present and hence the Green function is zero. This is just the requirement of causality: no signal reaches the observer for $t < x - \bar x$. On the other hand, for $t>x-\bar x$ the contour picks up an infinite number of residues of the Gamma function in Eq.~\eqref{eq:A-} at the QNM frequencies and hence
\begin{widetext}
    \begin{align}
        G(x, \bar x, t) &= G_{\mathrm{QNM}} (x, \bar x, t) \; , \quad \text{for} \, t>x-\bar x \; .  \nonumber \\
         G_{\mathrm{QNM}} (x, \bar x, t) &= \mathrm{Re} \sum_{n \geq 0} \frac{(-1)^{n+1}}{ n!} \frac{\Gamma \big( 1-2\lambda-n\big)}{\Gamma \big( 1 - \lambda - n \big)^2} e^{\alpha(n+\lambda) (x - \bar x - t)} F \Big( \lambda, \bar \lambda , 1 - n - \lambda, \big(1+e^{-2 \alpha \bar x} \big)^{-1} \Big) \times \nonumber \\
         &\times F \Big( \lambda , \bar \lambda , 1 - n - \lambda, e^{-2 \alpha x} \big(1+e^{-2 \alpha x} \big)^{-1} \Big)  \label{eq:GQNM} \; .
    \end{align}
\end{widetext}

Hence, the Green function consists only in a QNM part, oscillating at the QNM frequencies~\eqref{eq:QNMfreq}. This is consistent with our physical expectation: for a perturbation situated to the left of the maximum of the potential, the signal has to travel through the maximum before reaching the observer, interacting with the potential and producing QNMs. Notice that this expression is the same than the one presented in~\cite{Casals:2009zh}.

\subsection{Second case: $\bar x > 0$}

When $\bar x>0$, the asymptotics of the hypergeometric function that we used previously are not valid any more. Instead, we use the decomposition of $\psi_H^-$ into wavefunctions properly normalized at infinity in Eq.~\eqref{eq:defConnectionCoeffs} to obtain the formula
\begin{align} \label{eq:GreenDecompositionInfinity}
    \tilde G(x, \bar x, \omega) &= \frac{1}{2 i \omega} \psi_\infty^-(\bar x, \omega) \psi_\infty^+( x, \omega) \nonumber \\
    &+ \frac{A^+}{2 i \omega A^-} \psi_\infty^+(\bar x, \omega) \psi_\infty^+( x, \omega) \; .
\end{align}
This decomposition will be crucial in the following, so let us pause a moment to describe it. First, let us remark that using the same properties of the hypergeometric function described in Section~\ref{sec:xbar<0} and the expression for $A^+$ in~\eqref{eq:A+}, the asymptotic form of $\tilde G$ for large $|\omega|$ is
\begin{align}
    \omega \tilde G(x, \bar x, \omega) e^{-i \omega t} &\underrel{|\omega| \rightarrow \infty}{\simeq}  e^{i \omega (x - \bar x - t)} \nonumber \\
    &+  e^{i \omega (x + \bar x - t)} e^{- \mathrm{Sgn(Re} \omega) \pi \omega/\alpha} \; , \label{eq:AsymptoticGreenBarx>0}
\end{align}
where $\mathrm{Sgn}$ denotes the sign function. Hence, we have decomposed the Green function into two components: a ``direct wave" $e^{i \omega (x - \bar x - t)}$ which travels directly from the perturbation at $\bar x$ to the observer, and a ``scattered wave" $ e^{i \omega (x + \bar x - t)}$ which first travels to the left of the perturbation at $\bar x$, reaches the maximum of the potential at $x=0$, and then is scattered back to the observer. These two terms need to be handled separately for the arc contribution to the integral: the first one has a transition from upper to lower half-plane at $t= x-\bar x $, and the second one at $t = x + \bar x$. 

The second important property of the decomposition in~\eqref{eq:GreenDecompositionInfinity} is that the analytic structure of the Green function becomes much more intricate. Indeed, from their definition in Eq.~\eqref{eq:defHomogeneousSolutions}, it appears that the homogeneous solutions $\psi_\infty^-$ and $\psi_\infty^+$ have poles for $\omega = i \alpha (k+1)$ and $\omega = - i \alpha (k+1)$ respectively, with $k \in \mathbb{N}$. Hence, the first line in the decomposition~\eqref{eq:GreenDecompositionInfinity} has poles along the whole imaginary axis, including $\omega = 0$. These poles are \textit{exactly} cancelled by the term on the second line, so that the resulting Green function has no remaining poles on the imaginary axis. 

Let us now describe how to close the contour integral defining the temporal Green function in Eq.~\eqref{eq:InverseLaplaceGree}, illustrated in Figure~\ref{fig:contour}. For $t<x-\bar x$, we simply close the contour on the upper half-plane. The arc contribution is zero using Eq.~\eqref{eq:AsymptoticGreenBarx>0} and Jordan lemma. This gives again $G=0$, consistent with the requirement of causality.

For early times $x-\bar x < t < x+\bar x$, we use the decomposition~\eqref{eq:GreenDecompositionInfinity} and close the contour for the first term on the lower half-plane, while we keep closing in the upper half-plane for the second term. This way, the contour picks residues on the imaginary axis both in the upper and lower half-planes. Using properties of the hypergeometric function when its third argument is a negative integer~\cite{NIST:DLMF}, we obtain the \textit{early times Green function} $G_\mathrm{E}$
\begin{widetext}
    \begin{align}
        G(x, \bar x, t) &=  G_\mathrm{E}(x, \bar x,t) \, , \quad \text{for} \, x-\bar x < t < x+\bar x \; ,  \nonumber \\
         G_{\mathrm{E}} (x, \bar x, t) &= \frac{1}{2} \sum_{k \geq 0} s_k \frac{(-1)^{k+1}}{(k!)^2} \Bigg| \frac{\Gamma(\lambda + k)}{\Gamma(\lambda)} \Bigg|^2 \big( e^{-\alpha k(x+\bar x + t)}  + e^{\alpha k(t-x-\bar x)} \big)  \nonumber \\
         &\times F \big(\lambda, \bar \lambda, k+1, e^{-2 \alpha x} \big(1+e^{-2 \alpha x} \big)^{-1} \big) F \big(\lambda, \bar \lambda, k+1, e^{-2 \alpha \bar  x} \big(1+e^{-2 \alpha \bar x} \big)^{-1} \big) \label{eq:GI} \; ,
    \end{align}
\end{widetext}
where $s_0 = 1/2$ and $s_{k \geq 1}=1$ otherwise. Several remarks are in order: first, notice that $G_\mathrm{E}$ contains a constant piece corresponding to the pole at $\omega = 0$, giving a permanent displacement of the field after the first signal crossed the observer. This constant piece is qualitatively very similar to the Green function for the wave operator in one dimension $-\partial_t^2+\partial_x^2$, which is
\begin{equation} \label{eq:G1D}
    G_{1D} = - 1/2 \Theta(t - x + \bar x)    \; .
\end{equation}
Indeed, notice that for $x \gg 1$ and $\bar x \gg 1$, the constant piece in $G_E$ is exactly $G_{1D}$ for  $x-\bar x < t < x+\bar x$. This is consistent with our physical intuition that the P\"{o}schl-Teller potential goes very quickly to zero far from its maximum. Notice that $G_{1D}$ was also obtained in the delta-function potential considered in Ref.~\cite{Lagos_2023,Chavda:2024awq}, and in the large-frequency approximation of a scalar waveform in~\cite{PhysRevD.55.468}. 

Second, we see that $G_\mathrm{E}$ contains both a decaying term $e^{-\alpha k(x+\bar x + t)}$, quite negligible for large $x$, and a growing term $ e^{\alpha k(t-x-\bar x)}$ which reaches its maximum when the QNM ringing starts at $t = x + \bar x$. This exponentially growing term smoothly connects the pre-QNM regime to the start of the ringdown. It is similar in nature to the additional modes mentioned in Ref.~\cite{Szpak:2004sf}, and is not found in the delta-function potential Green function of Ref.~\cite{Lagos_2023,Chavda:2024awq}. Notice that the characteristic time scale of exponential growth, $1/\alpha$, is equal to half of the decay time of the fundamental QNM.

Finally, once the perturbation reached the top of the potential i.e. for $t>x + \bar x$, we can close both contours on the lower half-plane. Computing the residues at the QNM poles (the only poles present in the Green function once we sum the two terms in~\eqref{eq:GreenDecompositionInfinity}), we find the same expression for $G$ than in Eq.~\eqref{eq:GQNM}. 

\subsection{Summary and comparison with numerical results} \label{sec:summary_numerical}

Collecting the results of the two preceding sections, we finally find that the complete Green function is (whatever the sign of $\bar x$): 
\begin{align}
    G(x, \bar x, t) &= \Theta(t-x+\bar x) \Theta(x+\bar x-t) G_\mathrm{E}(x, \bar x, t) \nonumber \\
    &+ \Theta(t-x-|\bar x|) G_\mathrm{QNM}(x, \bar x, t) \label{eq:totGreen} \; ,
\end{align}
where $G_\mathrm{E}$ is given in Eq.~\eqref{eq:GI} and $G_\mathrm{QNM}$ is given in Eq.~\eqref{eq:GQNM}.

\begin{figure}
    \centering
    \includegraphics[width=\columnwidth]{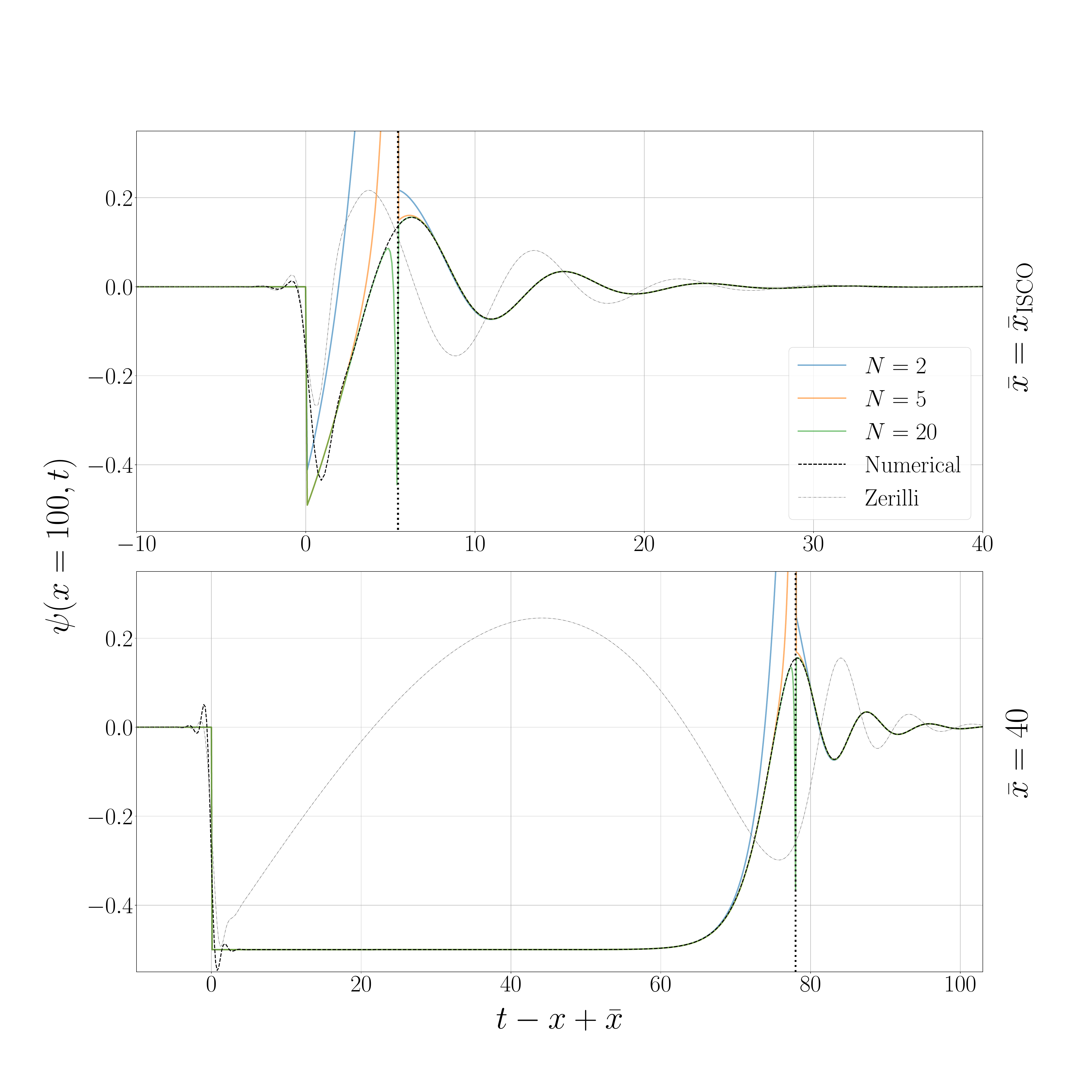}
    \caption{Comparison of the analytical Green function~\eqref{eq:totGreen} (shifted to ensure that the maximum of the potential is at $x = x_m$) and the numerical results for an observer situated at $x=100$. The upper panel shows the results for an initial perturbation at $\bar x = \bar x_\mathrm{ISCO} \simeq 3.69$ at the ISCO, while the lower panel has $\bar x = 40$. The continuous lines show the analytical results with an increasing number of terms $N$ in the sums~(\ref{eq:GQNM},\ref{eq:GI}). The black dashed line shows the result of the numerical integration with the P\"{o}schl-Teller potential, while the gray dot-dashed line shows the numerical result for the Zerilli potential. A vertical line signals the QNM starting time at $t=x + |\bar x|$. As expected, the Zerilli Green function is quite different from the P\"{o}schl-Teller one for $\bar x = 40$, because at large distances the two potentials are different (Zerilli has power-law tails, while P\"{o}schl-Teller does not). Notice also that the numerical waveform displays unphysical oscillations around $t=x-\bar x$ due to the finite width of the Gaussian used to model the delta-function. }
    \label{fig:GF_comparison}
\end{figure}

To test the validity of our analytical formula, we now compare it against numerical integration of the homogenous part of the differential equation~\eqref{eq:RWZeq} with the P\"{o}schl-Teller potential. Using Eq.~\eqref{eq:LeaverStrangeFormula}, we see that $\psi(x,t) = G(x, \bar x, t)$ if we specify the initial conditions at $t=0$:
\begin{equation} \label{eq:BCGreen}
    \psi(x,t=0) = 0 \; , \quad \frac{\partial \psi}{\partial t} \bigg|_{x,t=0} = - \delta(x-\bar x) \; .
\end{equation}
We wrote a simple code to numerically solve the differential equation in time-domain~\eqref{eq:RWZeq} using the fourth-order Runge-Kutta scheme. The function $\psi$ is evaluated on a discretized grid, using a fourth-order stencil to compute spatial derivatives. Spatial boundary conditions are ingoing at the left and outgoing at the right ($\partial_t \psi \pm \partial_x \psi = 0$), with ghost points (needed to compute the second order spatial derivative) ensuring that the boundary conditions are obeyed. The delta functions are mollified using a gaussian $\delta(x-\bar x) \rightarrow  (2 \pi \sigma^2)^{-1/2} e^{-(x-\bar x)^2/(2 \sigma^2)} $, with $\sigma = 0.8$. 

In Figure~\ref{fig:GF_comparison} we show the comparison of the numerical integration with initial data given in~\eqref{eq:BCGreen} and our analytical prediction for two kind of perturbations $\bar x$ situated at the Innermost Stable Circular Orbit (ISCO) of \sch spacetime ($\bar x = 3.69)$ and far from the maximum of the potential ($\bar x = 40$). 
We observe a nearly perfect agreement between the numerical and analytical result when the number of terms in the sums~(\ref{eq:GQNM},\ref{eq:GI}) are large enough. However, we also learn an important fact: while the QNM sum in Eq.~\eqref{eq:GQNM} converges quickly at the ringdown start time $t=x + |\bar x|$, the additional mode sum in the early times Green function~\eqref{eq:GI} converges very slowly at that precise time. This can be seen from Figure~\ref{fig:GF_comparison} as a divergence in the analytical Green function for times slightly before $t=x + |\bar x|$, while there is no such divergence slightly after $t=x + |\bar x|$. The divergence is confined to be closer and closer to the ringdown start time if we increase the number of terms in the sum~\eqref{eq:GI}. From the analytic formula~\eqref{eq:GI}, one can check that for large $x$ and $\bar x$ the sum converges like $\sum_k (-1)^k/k$, so quite slowly. This convergence does not seem to change much if we vary $\bar x$.
This unfortunately means that the additional exponentially growing modes that we found in Eq.~\eqref{eq:GI} do not enjoy the universality of QNM frequencies: fitting a waveform by adding a few of these modes is not doable in practice. 

Notice that Ref.~\cite{Szpak:2004sf} found very similar additional modes emerging before the ringdown start time for an exponential potential, however in this case the sum was quickly convergent and the Green function was approximated by the lowest-lying mode. It would be interesting to understand the origin of this discrepancy and more generally which class of potentials have an additional mode sum that is quickly convergent, but this is out of scope of this work. We now turn to the task of computing the waveform of a point-particle plunging into a BH using the analytical formula~\eqref{eq:totGreen} for the Green function.


\section{Response to an infalling particle} \label{sec:response}

Now that we have obtained a closed-form expression for the Green function in~\eqref{eq:totGreen}, we turn to the problem of finding the field generated by a point-particle infalling into the BH. We will consider two different trajectories of the point-particle: a purely radial infall, and the geodesic trajectory obtained from an initially circular orbit placed at the Innermost Stable Circular Orbit (ISCO) of the BH at $r=3$ (the ``universal plunge''). Both of these trajectories are described in Appendix~\ref{app:trajectories_source}, and they provide a relation $\bar t_\mathrm{P}(\bar x)$ (or equivalently $\bar x_\mathrm{P}(\bar t)$) between the coordinate time and distance of the particle, where $\bar x_P$ decreases with time. We also denote by $\bar v(\bar t) = \dd \bar x_\mathrm{P}/\dd \bar t<0$ the coordinate velocity of the particle.
We assume that we can initialize our simulation sufficiently far in the past so the we can neglect the impact of initial conditions, hence setting $\psi_0(x) = \dot \psi_0(x) = 0$. Thus, using Eq.~\eqref{eq:LeaverStrangeFormula} we can get the field $\psi(x,t)$ from
\begin{equation}
    \psi(x,t) = \int_{- \infty}^\infty \mathrm{d}\bar x \int_{t_0}^\infty \dd \bar t \; G(x-x_m, \bar x-x_m, t-\bar t) S(\bar x, \bar t) \; ,
\end{equation}
where we have taken into account the fact that the Green function that we obtained in Section~\ref{sec:PTGreen} included a linear shift of the tortoise coordinate to place the maximum at $x=0$, and where $S$ is the source term representing the point-particle, which takes the following form:
\begin{equation} \label{eq:Sourcefg}
    S(\bar x, \bar t) = f(\bar x_\mathrm{P}(\bar t)) \delta(\bar x - \bar x_\mathrm{P}(\bar t)) + g(\bar x_\mathrm{P}(\bar t)) \delta'(\bar x - \bar x_\mathrm{P}(\bar t)) \; ,
\end{equation}
where the expression of the functions $f$ and $g$ are given in Appendix~\ref{app:trajectories_source}. Plugging in the expression for the Green function that we found in~\eqref{eq:totGreen}, we obtain that the waveform can be decomposed in an \textit{instantaneous} (or impulsive) contribution $\psi_\mathrm{Inst.}$, travelling exactly on the light-cones of the Green function, and a \textit{historical} (or activation) piece $\psi_\mathrm{Hist.}$, depending of the past trajectory of the particle inside the light-cone. These two contributions take the form
\begin{widetext}
    \begin{align}
        \psi_\mathrm{Inst.} (x,t) &= \frac{g(\bar x_\mathrm{P}(\bar t))}{1+\bar v(\bar t) \, \mathrm{Sgn}(\bar x - x_m)} \bigg( \mathrm{Sgn}(\bar x - x_m) G_\mathrm{QNM}  - \Theta(\bar x-x_m) G_E \bigg) \bigg|_{\bar t = \bar t_\mathrm{QNM}(u), \bar x = \bar x_\mathrm{QNM}(u)} \nonumber \\
        &- \frac{g(\bar x_\mathrm{P}(\bar t))}{1-\bar v(\bar t) } \Theta(\bar x-x_m) G_E \bigg|_{\bar t = \bar t_D(u), \bar x = \bar x_D(u)} \; , \\
        \psi_\mathrm{Hist.}(x,t) &= \int_{t_0}^{t_\mathrm{QNM}(u)} \big[ f(\bar x_\mathrm{P}(\bar t)) G_\mathrm{QNM}  - g(\bar x_\mathrm{P}(\bar t)) \partial_{\bar x}  G_\mathrm{QNM}  \big] \big|_{\bar x = \bar x_P(\bar t)} \dd \bar t \nonumber \\
        &+ \int_{t_\mathrm{QNM}(u)}^{t_\mathrm{D}(u)} \big[f(\bar x_\mathrm{P}(\bar t)) G_\mathrm{E} - g(\bar x_\mathrm{P}(\bar t)) \partial_{\bar x}  G_\mathrm{E} \big] \big|_{\bar x = \bar x_P(\bar t)} \dd \bar t  \label{eq:psiHist} \; .
    \end{align}
\end{widetext}
In order to avoid clutter, we have suppressed the arguments of the Green functions, but they are always evaluated at $G(x-x_m, \bar x-x_m, t-\bar t)$. 
Furthermore we have defined $u=t-x+x_m$ the retarded time of the observer, and the retarded times $\bar t_\mathrm{QNM}(u)$ and $\bar t_\mathrm{D}(u)$ correspond to propagation along the QNM and direct signal light-cones respectively. They solve the equations:
\begin{align} \label{eq:QNMLightCone}
    \bar t_\mathrm{QNM} + |\bar x_P(\bar t_\mathrm{QNM}) - x_m| &= u \; , \\
    \bar t_\mathrm{D} - \bar x_P(\bar t_\mathrm{D}) + x_m &= u \; .
\end{align}
Physically, $\bar t_D$ corresponds to the time at which was emitted a signal travelling directly from the point-particle and reaching the observer at $t$, while $\bar t_\mathrm{QNM}$ is the time at which was emitted a signal travelling from the point-particle, interacting with the maximum of the potential at $x=x_m$, and then reaching the observer at $t$. Since the total distance that the signal travels for $\bar t_\mathrm{QNM}$ is longer than for $\bar t_D$, it has to start earlier in order to reach the observer at $t$, hence implying $\bar t_D \geq \bar t_\mathrm{QNM}$ (with $\bar t_D = \bar t_\mathrm{QNM}$ for $\bar x_P<x_m$). Equivalently, we can define $\bar x_\mathrm{QNM}$ and $\bar x_\mathrm{D}$, and they will satisfy $\bar x_D \leq \bar x_\mathrm{QNM}$. Notice that, as emphasized in the Introduction, these times do not necessarily correspond to propagation along a \sch geodesic since we are focussing here on a single $\ell$ mode.

Several remarks are in order. First, we have assumed $t_\mathrm{QNM}\geq t_0$, i.e. initial conditions are sufficiently far in the past so that the observer sees a signal that already interacted with the maximum of the potential. Second, notice that the piece involving $G_E$ in $\psi_\mathrm{hist}$ vanishes after the observer sees the point-particle crossing the maximum of the potential, since $\bar t_\mathrm{D} = \bar t_\mathrm{QNM}$ after that point. This is consistent with our finding in Section~\ref{sec:PTGreen} that a signal emitted to the left of the maximum only contains a QNM part. Finally, we can simplify further the expression of $\psi_\mathrm{Inst.}$ using the fact that the Green function is continuous at the ringdown start time $\bar t = \bar t_\mathrm{QNM}$ for perturbations outside the maximum $\bar x - x_m >0$~\footnote{This property is formally true only summing an infinite number of terms in $G_E$~\eqref{eq:GI}, as the sum otherwise diverges at that time. }. This gives $G_E(x-x_m, \bar x_\mathrm{QNM} - x_m, t-\bar t_\mathrm{QNM}) = G_\mathrm{QNM}(x-x_m, \bar x_\mathrm{QNM}-x_m, t-\bar t_\mathrm{QNM})$ for $\bar x_\mathrm{QNM} - x_m >0$. Hence
\begin{align} \label{eq:psiInst}
    &\psi_\mathrm{Inst.}(t,x) = - \frac{g(\bar x_\mathrm{P}(\bar t))}{1-\bar v(\bar t) } \Theta(\bar x-x_m) G_E \bigg|_{\bar t = \bar t_D(u), \bar x = \bar x_D(u)} \nonumber \\
    &-  \frac{g(\bar x_\mathrm{P}(\bar t))}{1-\bar v(\bar t) } \Theta(x_m - \bar x) G_\mathrm{QNM} \bigg|_{\bar t = \bar t_\mathrm{QNM}(u), \bar x = \bar x_\mathrm{QNM}(u)} \; .
\end{align}
This has the following physical interpretation: when the particle is situated to the right of the maximum, the observer only receives an instantaneous signal propagating directly from the particle and not interacting with the maximum (hence the use of $G_E$), while when the particle is to the left of the maximum, the signal always interact with the maximum (hence the use of $G_\mathrm{QNM}$). 

Notice that this last property implies that the impulsive contribution to the waveform $\psi_\mathrm{Inst.}$ computed in~\cite{DeAmicis:2025xuh} is probably inaccurate before the particle crosses the maximum of the potential. This is because in this study the authors consider only the QNM part of the Green function while neglecting the other pieces. As we have just seen, the early times Green function just gives a contribution equal in magnitude to the QNM part in $\psi_\mathrm{Inst.}$ before the particle crosses the maximum.

\begin{figure}
    \centering
    \includegraphics[width=1\columnwidth]{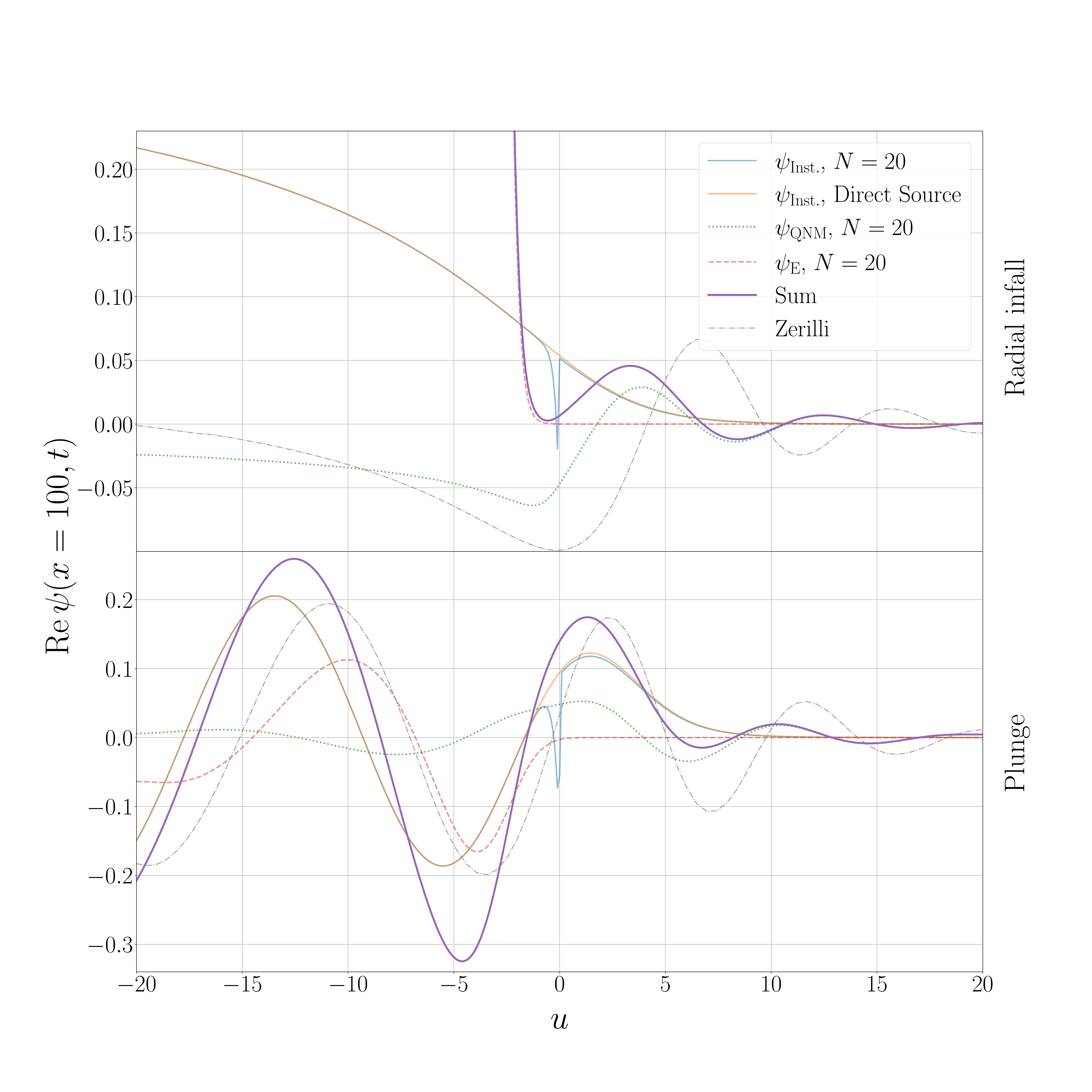}
    \caption{Real part of the wavefunction $\psi$ for a radial infall and a plunge trajectory. We plot each component of the waveform in Eqs.~(\ref{eq:psiInst},\ref{eq:psiHist}) separately, where $\psi_\mathrm{QNM}$ refers to the first line of Eq.~\eqref{eq:psiHist} and $\psi_\mathrm{E}$ to the second line of Eq.~\eqref{eq:psiHist}. We also plot the sum of these three waveforms and the numerical solution to the Zerilli equation using the same trajectory and the code described in Section~\ref{sec:summary_numerical}. 
    $N=20$ means the wavefunction obtained by summing 20 terms in Eqs.~(\ref{eq:GQNM},\ref{eq:GI}), while ``Direct Source" refers to the simplified expression~\eqref{eq:psiInstSimplified}. The divergence of $\psi_\mathrm{Inst.}$ close to $u=0$ is the same than the one discussed below Figure~\ref{fig:GF_comparison}.}
    \label{fig:psiTot}
\end{figure}

We plot (the real part of) the waveform as observed by a distant observer situated at $x=100$ in Figure~\ref{fig:psiTot}, as a function of the retarded time $u=t-x+x_m$ and for the two point-particle trajectories described in Appendix~\ref{app:trajectories_source}. We have normalized the point-particle time (see Appendix~\ref{app:trajectories_source}) such that $u=0$ when the point-particle is situated at the maximum of the potential $x_P=x_m$. We now discuss each part of the waveform separately, highlighting their main features and limits.

\subsection{Instantaneous part}


From Figure~\ref{fig:psiTot} we can observe that $\psi_\mathrm{Inst.}$ is nonzero for $u<0$ but very quickly goes to zero after the point-particle crosses the maximum of the potential. 
A surprising feature that comes out from Figure~\ref{fig:psiTot} is that $\psi_\mathrm{Inst.}$ is very well approximated if one just set $G_\mathrm{E} = G_\mathrm{QNM} = -1/2$ in Eq.~\eqref{eq:psiInst}, i.e.:
\begin{align} \label{eq:psiInstSimplified}
    &\psi_\mathrm{Inst.}(t,x) \simeq  \frac{g(\bar x_\mathrm{P}(\bar t))}{2(1-\bar v(\bar t)) }  \bigg|_{\bar t = \bar t_D(u), \bar x = \bar x_D(u)} \; .
\end{align}
Why is it so? Concerning $G_E$, this can be interpreted from the fact that $G_E \simeq -1/2$ at the time of the first signal $\bar t = \bar t_D(u)$ when $\bar x \gg 1$ (see e.g. the bottom panel of Fig.~\ref{fig:GF_comparison} and the comment concerning $G_{1D}$ in Eq.~\eqref{eq:G1D}). Concerning $G_\mathrm{QNM}$, one can observe that at the onset of ringdown $\bar t = \bar t_\mathrm{QNM}(u)$ and for $x \gg 1$, $|\bar x| \gg 1$, $\bar x <0$ one has from Eq.~\eqref{eq:GQNM}
\begin{align}
    &G_{\mathrm{QNM}} (x-x_m, \bar x_\mathrm{QNM}-x_m, t-\bar t_\mathrm{QNM})  \nonumber \\
    &\simeq \mathrm{Re} \sum_{n \geq 0} \frac{(-1)^{n+1}}{ n!} \frac{\Gamma \big( 1-2\lambda-n\big)}{\Gamma \big( 1 - \lambda - n \big)^2} = - \frac{1}{2} \; .
\end{align}

Hence, the instantaneous part $\psi_\mathrm{Inst}$ just follows the source function $g(\bar x_P(\bar t))$. As can be seen from the expression of $g$ in Eq.~\eqref{eq:gfunc}, $\psi_\mathrm{Inst}$ is thus quickly redshifted away by the factor $f(r)$ once the particle approaches the horizon. On the other hand, at large radius one has $g \simeq \mathrm{Const.} \times e^{-i m \phi}$. This early time asymptotic can easily be observed in Figure~\ref{fig:psiTot}: this is a constant for the radial infall (implying a kind of memory effect), and an oscillating function set by the frequency of the point-particle for the universal plunge. 

\subsection{Historical part from $G_\mathrm{QNM}$} \label{sec:historicalQNM}


In Figure~\ref{fig:psiTot} we plot the part of $\psi_\mathrm{Hist}$ coming from the QNM Green function, i.e. the first line of Eq.~\eqref{eq:psiHist}. In both cases it starts smoothly from zero at early times, reaches a peak when the particle crosses the maximum of the potential, and then display oscillations at the QNM frequencies. In the universal plunge case, one can also easily observe an oscillation set by the point-particle frequency before $u=0$, followed by QNM ringing after $u=0$. Hence, the QNM wavefunction can contain frequencies different from the QNM values before $u=0$.

Let us comment on the asymptotics of the QNM waveform for times well after $u=0$. The standard picture is that the waveform should contain an infinite number of QNM frequencies, labeled by their overtone number $n$. However, as noted in Ref.~\cite{DeAmicis:2025xuh}, overtones for $n \geq 1$ are in fact swamped by additional modes which enter the waveform, the so-called \textit{redshift modes}. In our toy-model, we indeed recover the results of Ref.~\cite{DeAmicis:2025xuh}, as we will now show.

When the particle is approaching the horizon, its trajectory becomes universal with the following relation between $t_P$ and $x_P$:
\begin{equation}
    \bar t_P(\bar x_P) \underrel{x_p \rightarrow \infty}{\simeq} - \bar x_P \; .
\end{equation}
This implies that a signal travelling on the QNM light cone~\eqref{eq:QNMLightCone} has $t_P = u/2$. The QNM Green function in Eq.~\eqref{eq:GQNM} hence becomes
\begin{align}
    &G(x-x_m, \bar x-x_m, t-\bar t) \nonumber \\
    & \simeq \sum_n \gamma_n  e^{-\alpha(n + \lambda) u} e^{2 \alpha (n+\lambda) \bar t} \; , \text{for} \, \bar t \simeq \bar t_P, \bar x \simeq \bar x_P \; ,
\end{align}
where $\gamma_n$ are constant coefficients. On the other hand, the functions $f$ and $g$ behave like $e^{x}$ for $x \rightarrow - \infty$ (see Appendix~\ref{app:trajectories_source}). This means that the integral defining $\psi_\mathrm{Hist}$ in Eq.~\eqref{eq:psiHist} scales as
\begin{align}
    \psi_\mathrm{Hist}\simeq  \sum_n \gamma_n e^{-\alpha(n + \lambda) u}  \int^{\bar t_P} e^{(2 \alpha(n+\lambda) - 1)\bar t } \dd \bar t \; .
\end{align}
The behavior of $\psi_\mathrm{Hist}$ depends on whether the integral is convergent. For $n=0$, one has $\mathrm{Re}(2 \alpha \lambda-1)<0$, so the integral converges and we have the expected QNM oscillation at the frequency $\omega_n = - i n (\alpha + \lambda)$. On the other hand, for $n \geq 1$ one has $\mathrm{Re}(2 \alpha (n+\lambda)-1)>0$ and the integral diverges. This divergence (which in fact was thrown away in previous computations of QNM amplitudes~\cite{Leaver:1986gd,Hadar:2009ip,Zhang:2013ksa}) exactly cancels the QNM oscillation to leave an exponentially damped term,
\begin{equation}
    \psi_\mathrm{Hist} \propto e^{-u/2} \; .
\end{equation}
This damping corresponds to the surface gravity of the BH, and these kinds of modes were already identified in other works~\cite{Mino:2008at,Zimmerman:2011dx,Oshita:2025qmn}. In fact, the instantaneous part of the waveform $\psi_\mathrm{Inst.}$ is also a pure redshift mode as the particle approaches the horizon, as is clear from Eq.~\eqref{eq:psiInstSimplified} and the asymptotics of the source function $g$. 
Notice that it has been argued in~\cite{Oshita:2025qmn} that these horizon modes are in fact screened due to a zero of the Green function at this frequency: we disagree with such a statement, and we will elaborate on this point in Section~\ref{sec:remarks}.


\subsection{Historical part from $G_\mathrm{E}$}


The historical part from $G_E$, i.e. the second line of Eq.~\eqref{eq:psiHist}, is finally plotted in the same Figure~\ref{fig:psiTot}. As highlighted after Eq.~\eqref{eq:psiHist}, it vanishes after $u=0$. 
The radial infall is perhaps the most surprising result: for negative times the waveform quickly grows without bound, showing a divergence. This unphysical feature can be understood from the fact that, in the far past, one can approximate $G_E \simeq -1/2$ so that the historical part from $G_E$ becomes
\begin{equation}
    \psi_\mathrm{Hist.} \simeq - \frac{1}{2}  \int_{t_\mathrm{QNM}(u)}^{t_\mathrm{D}(u)} f(\bar x_\mathrm{P}(\bar t)) \dd \bar t \; .
\end{equation}
When $u \rightarrow - \infty$, the difference between $t_\mathrm{QNM}$ and $t_D$ is large, hence the divergence in $\psi_\mathrm{Hist.}$. For the radial plunge, one can indeed check that $\psi_\mathrm{Hist.} \simeq \mathrm{Const.} \times |u|^{1/3}$ for large negative $u$. For the radial infall instead, the oscillations in $f(x_P)$ make the integral going to zero.

Of course, such a divergence does not happen when solving the true RWZ equations. This can be heuristically understood from the fact that, at large distances, the stronger scattering from the power-law tail in the RWZ potentials produce negative interference which result in a vanishing wavefunction. Indeed, note that the Green function of the true RWZ equations, plotted in Figure~\ref{fig:GF_comparison}, is not a constant $-1/2$ term at times between the direct signal and QNM ringing, but instead displays nontrivial oscillations.

\section{Remarks on the analytic structure of the RWZ equations} \label{sec:remarks}

Our results, even if it is clear from Figure~\ref{fig:psiTot} that they do not fit so well the true solution to the RWZ equations or even display an unphysical divergence, have still allowed us to better understand how to compute the Green function before the QNM starting time. Can we apply the same technique to the full problem, replacing the P\"{o}schl-Teller potential with the RWZ expressions? We will see that the branch cut of the Green function at $\omega=0$ makes the problem more complicated, but it is still interesting to understand the analytic structure of the Green function, which is the topic of this Section.

The Green function, as defined in Eq.~\eqref{eq:defGreen}, requires the solutions to the homogeneous RWZ equations. For definiteness, let us concentrate on the Regge-Wheeler equation, for which $V(r) = (1-1/r)(\ell(\ell+1)/r^2 - 3/r^3)$. By doing the change of variables $\tilde \psi(r, \omega) = r^3 e^{-i \omega (r+\log (r-1))} H(r, \omega) $ we find that the homogeneous Regge-Wheeler equation take the form of a Heun equation~\cite{Fiziev:2005ki},
\begin{align}
    &\frac{\dd H}{\dd r^2} + \bigg(-2i \omega + \frac{1-2 i \omega }{r-1} + \frac{5}{r} \bigg) \frac{\dd H}{\dd r} \nonumber \\
    &+ \frac{1}{r(r-1)} \big(6 - \ell(\ell+1) - 6 i \omega r \big) H(r) = 0 \; .
\end{align}
We denote by $H_H^-$ the solution to this equation normalized so that $H_H^-(1) = 1$ (i.e. $\tilde \psi = \psi_H^-$ is ingoing at the horizon). It is given in terms of the confluent Heun function by
\begin{equation}
    H_H^- = \mathrm{HeunC}( \ell(\ell+1)-6+6 i \omega, 6 i \omega, 1-2i\omega, 5, 2i \omega; 1-r) \; .
\end{equation}
The confluent Heun function is itself defined as an infinite series expansion around $1-r = 0$:
\begin{equation}
    H_H^-(r, \omega) = \sum_{n=0}^\infty b_n (1-r)^n \; ,
\end{equation}
where the coefficients $b_n$ obey the recurrence relation~\cite{2018arXiv180401007M}
\begin{align} \label{eq:RecurrenceHeun}
    P_n b_n &= Q_n b_{n-1} + R_n b_{n-2} \; , \\
    P_n &= n(-2 i \omega+n) , \nonumber \\
    Q_n &= 6 - \ell(\ell+1) - 6 i \omega + (n-1) (4-4i\omega +n) \; , \nonumber \\
    R_n &= 2 i \omega (n+1)  \nonumber \; ,
\end{align}
with $b_{-1} = 0$, $b_0 = 1$. This function is regular for $r>1$, and is analytic in $\omega$ apart from an ensemble of simple poles which are obvious from the recurrence relation~\eqref{eq:RecurrenceHeun}: these are the zeros of $P_n$, i.e. the set $\omega_n = - i n/2$ for an integer $n \geq 1$. 

On the other hand, we can also look for an asymptotic series expansion of $H$ for large values of $r$. In our specific case, this allows us to define $H_\infty^-$ (corresponding to a solution ingoing at infinity $\tilde \psi = \psi_\infty^-$) in the following way~\cite{2018arXiv180401007M}:
\begin{equation} \label{eq:HeunInfty}
    H_\infty^-(r, \omega) = r^{-3} \sum_{n=0}^\infty \beta_n \frac{(-1)^n n!}{(2i\omega r)^n} \; ,
\end{equation}
where this time the coefficients $\beta_n$ obey the recurrence relation
\begin{align}
    \beta_n &= Q_n \beta_{n-1}+R_n \beta_{n-2} \; , \\
    Q_n&= 1 - \frac{n+\ell(\ell+1)}{n^2} \nonumber \; , \\
    R_n &=  2i\omega \frac{(n+1)(n-3)}{n^2(n-1)} \nonumber \; ,
\end{align}
where $\beta_{-1}=0$, $\beta_0=1$. 
It is important to note that Eq.~\eqref{eq:HeunInfty} is never convergent for any value of $r$, so it should be understood in terms of asymptotic series, giving a good approximation to $ H_\infty^-(r, \omega)$ for $r$ large enough by optimal truncation. This time, $\psi_\infty^-$ does not have poles in the complex plane, but it has a branch point at $\omega=0$ which can be seen from the breakdown of the asymptotic series~\eqref{eq:HeunInfty} at that point. A naive and easy way to define $\psi_\infty^+$ is just to send $\omega \rightarrow - \omega$:
\begin{equation} \label{eq:rel_psi+_psi-}
    \psi_\infty^+(r, \omega) = \psi_\infty^-(r, -\omega) \; .
\end{equation}
The Wronskian $W(\omega) = \psi_H^- (\psi_\infty^+)' - (\psi_H^-)' \psi_\infty^+$ then inherits the analytic structure of both $\psi_H^-$ and $\psi_\infty^+$, i.e. an infinite series of simple poles on the negative imaginary axis and a branch point at $\omega=0$. 
Moving forward, the Green function, as defined in Eq.~\eqref{eq:defGreen}, does not have any pole on the imaginary axis because the poles of the Wronskian cancel the ones in $\psi_H$. We are hence in disagreement with a recent study in~\cite{Oshita:2025qmn} claiming that the zeros of $1/W$, i.e. the poles of the Wronskian, screen the horizon modes that we investigated in Section~\ref{sec:historicalQNM}. Indeed, the full Green function is regular at these values of $\omega$, with no zero nor pole, while the horizon modes just emerge as a property of the source term in the RWZ equation. The oversight of Ref.~\cite{Oshita:2025qmn} is to approximate $\psi_H^- \simeq e^{-i \omega r}$ close to the horizon, thus losing the pole structure of this function.  

On the other hand, there remains a branch point in the Green function at $\omega=0$ which is well-known and has been described by Leaver~\cite{Leaver:1986gd}. 
This branch point presents a technical issue which prevents us from applying the methods used in this article to the problem of finding the Green function of the RWZ equation. Indeed, let us imagine defining a branch cut for $\psi_\infty^+$ along the negative imaginary axis, as it is usually done in the literature. Then, the relation~\eqref{eq:rel_psi+_psi-} implies that -- with our naive definition -- $\psi_\infty^-$ has a branch cut along the \textit{positive} imaginary axis. A decomposition of the Green function as in Eq.~\eqref{eq:GreenDecompositionInfinity} is thus meaningless at face value, since it splits the Green function into two pieces which have a branch cut along the whole imaginary axis. There are other ways to define both $\psi_\infty^+$ and $\psi_\infty^-$ such that they have a branch cut on the same axis (these are multi-valued functions, so they can be defined on different branches), see for example the article from Leaver~\cite{Leaver:1986gd} for more details. 
Still, it is probable that the branch cut itself does determine somes of the properties of the early times Green function, so that it could contribute in a non-negligible way to the waveform at early times. We leave such a study to future work.

\section{Conclusions}

In this work we have used an approximation of the RWZ potential in order to analytically compute the time-domain Green function of BH perturbations, taking into account all causality conditions. One of our most interesting result is the existence of additional modes which are exponentially growing before the onset of ringdown, with a growing time for the lowest-order mode equal to half of the decay time of the fundamental QNM. Unfortunately, in order to accurately reproduce the Green function we needed to add many of these modes together, because the infinite sum containing them converges very slowly just before the ringdown start time. This means that these new modes are probably useless from a practical perspective, because they cannot improve a fit to the waveform with a few parameters only. Still, it would be interesting to understand if this additional mode sum is also present for the Green function of the true RWZ potential. This necessitates to better understand the properties of the branch cut and large frequency arcs integral, which determines the behavior of the early time piece of the Green function.

Another result worth of interest is our splitting~\eqref{eq:psiHist}-\eqref{eq:psiInst} of the waveform between an instantaneous part, just tracking the source of the RWZ equations on the light-cones of the Green function, and an historical part represented as an integral over the past history of the system. As expected, after the particle crosses the maximum of the potential all contributions go to zero quicker than the QNM part which dominate the signal. As the particle approaches the horizon, we have highlighted the presence of redshift modes both in the instantaneous and QNM part of the waveform. 
In our toy-model, the waveform can present unphysical divergences before merger which should be regulated by the power-law tail of the true RWZ potentials. Still, we expect such a decomposition to exist also for the RWZ Green function, so that our toy-model can give a reasonable physical intuition of what should happen for the realistic case of RWZ equations. In particular, we want to highlight the fact that the early times Green function contributes in an important way to the instantaneous part of the waveform. In order to compute this contribution to the RWZ Green function, we have to better understand the analytic structure of their homogeneous solutions and in particular of the branch cut, which is left to future work.

\section*{Acknowledgments}

I would like to thank Marina de Amicis, Enrico Cannizzaro, Nicola Franchini, Leonardo Gualtieri and Matteo della Rocca for stimulating discussions and useful comments on the draft. I'm particularly grateful to Laura Sberna and Matteo della Rocca for finding two typos in Section~\ref{sec:remarks}. I thank the Fundação para a Ciência e Tecnologia (FCT), Portugal, for the financial support to the Center for Astrophysics and Gravitation (CENTRA/IST/ULisboa) through grant No. UID/PRR/00099/2025 and grant No. UID/00099/2025, as well as to the FCT project ``Gravitational waves as a new probe of fundamental physics and astrophysics'' grant agreement 2023.07357.CEECIND/CP2830/CT0003.

\appendix

\section{Point-particle trajectories and source functions} \label{app:trajectories_source}

In this appendix we first describe the two plunge trajectories that we consider for the point-particle, which are geodesics of \sch spacetime. Both of them are in the $\theta = \pi/2$ plane and are characterized by an energy $E$ and angular momentum $L$, with a velocity 
\begin{equation}
    v = \frac{\dd x}{\dd t} = - \frac{1}{E} \sqrt{E^2 - A(r) \bigg( \frac{L^2}{r^2}+1 \bigg)} \; ,
\end{equation}
where $A(r) = 1-1/r$.
For a purely radial infall from infinity, $E=1$ and $L=0$  so that we have the following trajectory:
\begin{align}
    \phi &= 0 \; , \\
    t(r) &= -\frac{2}{3} r^{3/2} - 2 r^{1/2} + \log \bigg( \frac{r^{1/2}+1}{r^{1/2}-1} \bigg) + t_0 \; ,
\end{align}
In our setup, we adjust $t_0$ such that $t=0$ when the point-particle crosses the maximum of the potential at $x=x_m$.

On the other hand, for a trajectory starting circular at the ISCO one has $E = 2 \sqrt{2}/3$, $L=\sqrt{3}$ and~\cite{Hadar:2009ip}
\begin{align}
    \phi(r) &= \bigg( \frac{12r}{3-r} \bigg)^{1/2} \; , \\
    t(r) &= \frac{2r(1-12/r)}{\chi} - 22 \sqrt{2} \tan^{-1} \big( \sqrt{2} \chi \big) \nonumber \\
    &+ 2 \tanh^{-1} \chi + t_0 \; ,
\end{align}
where $\chi = (3/r-1)^{1/2}/\sqrt{2}$. 

The functions $f$ and $g$ multiplying the delta-functions in the expression~\eqref{eq:Sourcefg} are given by~\cite{Nagar:2006xv}
\begin{align}
    g(r) &= \frac{16 \pi Y_{\ell m}^* A(r)}{r E \mu (3 + r (\mu -2))} \big( L^2+r^2 \big) \label{eq:gfunc} \; , \\
    f(r) &= \frac{16 \pi Y_{\ell m}^* A(r)}{r E \mu (3 + r (\mu -2))} \bigg[ -2 i m L E v + \frac{5}{2} - \frac{r}{2}(\mu -2) \nonumber \\
    &+ \frac{L^2}{r^2} \big( 3 + r (m^2-\mu -2) \big) + \frac{L^2}{r^2(\mu -2)} \big( 3m^2-\mu -5 \big) \nonumber \\
    &+ \frac{r}{3+(\mu -2) r} \bigg( 6E^2 - A(r) (\mu -2) \bigg(1+\frac{L^2}{r^2} \bigg) \bigg)\bigg] \; ,
\end{align}
where $\mu = \ell(\ell+1)$. We have normalized the point-particle mass to one. 
In the main text, we use the source for the multipole $\ell = m = 2$.

\bibliography{refs}

\end{document}